\documentclass[floatfix,aps,prb,twocolumn,groupedaddress]{revtex4-2}

\usepackage{changes}
\usepackage{mathtools}
\usepackage{physics}
\usepackage[english]{babel}
\usepackage{amsfonts}
\usepackage{amsmath}
\usepackage{siunitx}
 \usepackage{tikz}
 \usepackage{xcolor}
\usepackage[colorlinks, citecolor={blue}, urlcolor={blue}, linkcolor={red!80!black}]{hyperref}

\usepackage{etoolbox}
\newcommand{\zerodisplayskips}{%
 \setlength{\abovedisplayskip}{5pt}%
 \setlength{\belowdisplayskip}{5pt}%
 \setlength{\abovedisplayshortskip}{5pt}%
 \setlength{\belowdisplayshortskip}{5pt}}
\appto{\normalsize}{\zerodisplayskips}
\appto{\small}{\zerodisplayskips}
\appto{\footnotesize}{\zerodisplayskips}
\newcommand\varpm{\mathbin{\vcenter{\hbox{%
  \oalign{\hfil$\scriptstyle+$\hfil\cr
          \noalign{\kern-.3ex}
          $\scriptscriptstyle({-})$\cr}%
}}}}
\newcommand\varmp{\mathbin{\vcenter{\hbox{%
  \oalign{$\scriptstyle({+})$\cr
          \noalign{\kern-.3ex}
          \hfil$\scriptscriptstyle-$\hfil\cr}%
}}}}

\begin{document}


\title{Resonator-mediated quantum gate between distant charge qubits}


\author{Florian Kayatz}
\author{Jonas Mielke}
\author{Guido Burkard}
\email[Author to whom any correspondence should be addressed.\\ e-mail: ]{guido.burkard@uni-konstanz.de}
\affiliation{Department of Physics, University of Konstanz, D-78457 Konstanz, Germany}


\date{\today}

\begin{abstract}
    Strong charge-photon coupling allows the coherent coupling of a charge qubit, realized by a single charge carrier (either an electron or a hole) in a double quantum dot, to photons of a microwave resonator. Here, we theoretically demonstrate that, in the dispersive regime, the photons can mediate both an $i\mathrm{SWAP}$ gate as well as a $\sqrt{i\mathrm{SWAP}}$ gate between two distant charge qubits. We provide a thorough discussion of the impact of the dominant noise sources, resonator damping and charge qubit dephasing on the average gate fidelity. Assuming a state-of-the art resonator decay rate and charge qubit dephasing rate, the predicted average gate fidelities are below 90\%. However, a decrease of the charge qubit dephasing rate by one order of magnitude is conjectured to result in gate fidelities surpassing 95\%. 
\end{abstract}


\maketitle

\section{\label{sec:introduction}Introduction}

The advanced production methods in the microelectronics industry, in particular for semiconductors such as Si, paved the way for electrostatically defined Si/SiGe quantum dots (QD) and arrays of such QDs \cite{zajac2015,zajac2016,sigillito2019,volk2019,lawrie2020} with excellent control over the individual QD potentials thereby demonstrating the high capability for scaling in this quantum computing hardware approach. There are different ways to implement a qubit using quantum dots. For an electron confined in a QD, the spin \cite{loss1998,Burkard2023} or the valley  \cite{boross2016,schoenfield2017,stockklauser2017} degree of freedom are utilized to operate a qubit. Likewise, the spin of a confined hole can define the qubit states \cite{hendrickx2020a}. On the other hand, combining two nearby QDs to a double quantum dot (DQD) opens up the possibility for further qubit implementations. The most basic among these is the charge qubit \cite{gorman2005,petersson2010,dovzhenko2011,shi2013,yang2019} obtained by filling a DQD with a single charge carrier, either an electron or a hole. The same single charge carrier configuration can also be used to realize the flopping mode spin qubit  \cite{benito2019,croot2020,mutter2021}. The addition of a second electron or hole to the DQD allows for various types of singlet-triplet qubits \cite{barthel2009,barthel2010,maune2012,takeda2020,jirovec2021,jirovec2022}. 

Regardless of the specific qubit realization, building a quantum computer requires excellent control over the individual qubits in the form of single-qubit gates and the possibility to couple two qubits via entangling two-qubit gates such as the prominent CNOT gate, the $i\mathrm{SWAP}$ gate, as well as the $\sqrt{i\text{SWAP}}$ gate. The availabiltiy of an entangling two-qubit gate is essential for quantum computing because in combination with single-qubit gates arbitrary multi-qubit operations can be performed \cite{divincenzo1995}.
The aforementioned two-qubit gates are effectively equivalent to the CNOT operation, as this operation can be constructed using a combination of single-qubit gates and either $i\mathrm{SWAP}$ gates \cite{Schuch2003} or  $\sqrt{i\mathrm{SWAP}}$ gates \cite{Burkard1999_2, blais2004}. The set of intrinsic two-qubit gates available in a specific device is determined by its physical properties. For example, systems with an $XY$-interaction can implement intrinsic $i\mathrm{SWAP}$ and $\sqrt{i\mathrm{SWAP}}$ gates \cite{Rasmussen2020}.

For the practical implementation of quantum computation and quantum error correction, long-range coupling between distant qubits is highly desirable. An ideal situation would be the availability of  ``all-to-all" connectivity \cite{corcoles2015,debnath2016}, i.e., an architecture with connectivity between any desired pair of qubits. The direct coupling between electrons, such as the capacitive coupling \cite{fujisawa2011,shulman2012,li2015,nichol2017,neyens2019,cayao2020} and exchange \cite{veldhorst2015,zajac2018,watson2018,huang2019,xue2019}, can be harnessed to build two-qubit gates in semiconductor devices. However, these interactions are typically short-ranged and do not allow the coupling of distant qubits. There are various approaches to couple distant qubits including the physical shuttling of qubits \cite{taylor2005,li2017,flentje2017,fujita2017,feng2018,zhao2018,mills2019,ginzel2020,krzywda2020,buonacorsi2020,vandiepen2021,krzywda2021a,yoneda2021,noiri2022a,boter2022}, an inter-qubit coupling mediated by a ``jellybean" quantum dot \cite{croot2018,malinowski2018,malinowski2019,seidler2022,wang2023,langrock2023}, and adiabatic passage protocols \cite{kandel2021,greentree2004,petrosyan2010,chancellor2012,srinivasa2007,farooq2015,ban2019,pico-cortes2019,gullans2020}. Another promising approach is the circuit quantum electrodynamics architecture (cQED), as considered in this paper, where a microwave resonator acts as a mediator between spatially separated qubits. 

Circuit quantum electrodynamics architectures were initially established in the field of superconducting qubits \cite{Devoret2013, blais2021}. However, many of these concepts are also applicable to other systems \cite{Xiang2013} including semiconductor-based qubits \cite{burkard2020}.
Strong charge-photon coupling and strong spin-photon coupling \cite{benito2017,mi2018,samkharadze2018,borjans2020,harvey-collard2022,yu2023} have been realized in circuit quantum electrodynamics with semiconductor devices in the past few years. Therefore, microwave resonators constitute such an intermediate system and have proven to implement a coupling between charge qubits separated by 42~$\mu\mathrm{m}$ in the resonant and dispersive charge-photon coupling regime \cite{vanwoerkom2018a} as well as the coupling between spins separated by more than four millimeters in the resonant \cite{borjans2020} and the dispersive \cite{harvey-collard2022} spin-photon coupling regime. This result suggests that resonator-mediated two-qubit gates are within reach. In particular, a resonator-mediated $i\mathrm{SWAP}$ gate between spin qubits was suggested \cite{benito2019a} and has recently been demonstrated \cite{dijkema2023}. 
The mechanism generating spin-photon coupling relies on spin-charge hybridization due to a magnetic field gradient in case of electrons and intrinsic spin-orbit coupling in case of holes, and thus couples spin and photons via the charge-photon dipole coupling \cite{benito2017,mi2018,benito2019a,yu2023}. Therefore, the spin coherence time suffers from the charge admixture, while the spin-photon coupling strength is significantly reduced compared to the charge-photon one \cite{benito2017,yu2023}. Working in the strong-coupling regime of circuit QED with spin qubits thus requires a balance between spin-charge hybridization and spin coherence.
Here, we study the simpler case of pure charge qubits that allow for strong coupling and short gate times at the expense of significantly lower coherence times.  Moreover, in case of electrons, the absence of micromagnets significantly simplifies the device design.

We theoretically demonstrate that microwave resonator photons can mediate both an $i\mathrm{SWAP}$ gate as well as a $\sqrt{i\mathrm{SWAP}}$ gate between distant charge qubits defined in a single charge carrier DQD and dispersively coupled to the resonator. To assess the potential of the gate for quantum computation,  
the effects of resonator photon decay and charge qubit dephasing on the gate fidelity are systematically discussed. Assuming a state-of-the art resonator decay rate and charge-qubit dephasing rate, the predicted average gate fidelities are below 90\%. However, an increase of the charge qubit dephasing rate by one order of magnitude is conjectured to result in gate fidelities surpassing 95\%.

\section{Model}

\begin{figure}[t]
    \centering
    \includegraphics[width=\columnwidth]{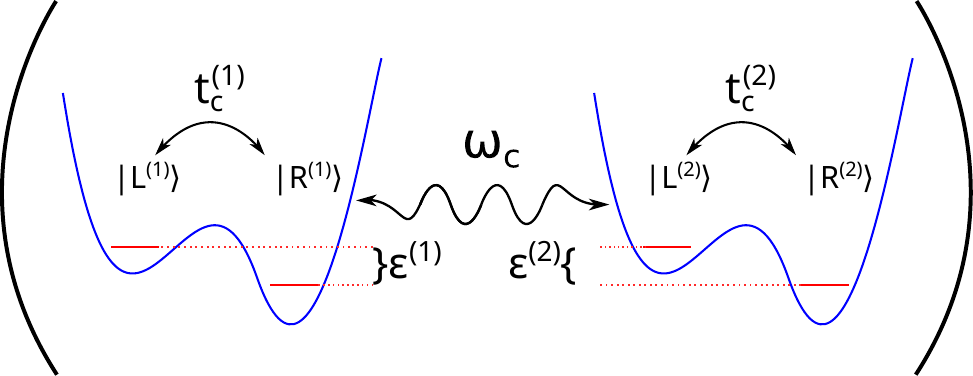}
    \caption{Schematic illustration of two ($i=1,2$) Si/SiGe DQDs coupled to a single mode ($\omega_c$) of a microwave resonator. Both DQDs are populated by a single charge carrier. In each case, the energy levels of the left (L) and right (R) QD are detuned by an amount $\varepsilon^{(i)}$ and the states $\vert L^{(i)}\rangle$ and $\vert R^{(i)}\rangle$ are tunnel coupled with strength $t_c^{(i)}$.}
    \label{fig:system}
\end{figure}

We consider two ($i=1,2$) Si/SiGe DQDs each populated with a single charge carrier (Fig.~\ref{fig:system}), either an electron or a hole. The respective charge carrier can either be confined to the $\ket{L^{(i)}(\text{eft})}$ or $\ket{R^{(i)}(\text{ight})}$ QD of the DQD. 
The energy difference between the levels of the two QDs is parameterized by the detuning parameter $\varepsilon^{(i)}$, while the QDs are tunnel-coupled with coupling strength $t_c^{(i)}$. Thus, the dynamics of the individual DQDs are described by the Hamiltonian
\begin{equation}
    \tilde{H}^{(i)}_{\mathrm{DQD}} = \frac{\varepsilon^{(i)}}{2}\tilde{\tau}_z^{(i)} + t_c^{(i)} \tilde{\tau}_x^{(i)}\,,
    \label{eq:HDQD}
\end{equation}
with the position space Pauli operators $\tilde{\tau}_j^{(i)}$ defined by $\tilde{\tau}_z^{(i)}\ket{L^{(i)}(R^{(i)})} =\varpm \ket{L^{(i)}(R^{(i)})}$. 

Being interested in long-ranged resonator-mediated two-qubit gates, we look at a setup where both DQDs are coupled to the same mode of a microwave resonator with frequency $\omega_c$. The Hamiltonian for the relevant resonator mode reads
\begin{equation}
H_c = \omega_c \left(a^\dagger a + \frac{1}{2}\right),
\end{equation}
with $a$ and  $a^\dagger$ the resonator photon annihilation and creation operators, respectively. Each single charge carrier DQD has a substantial electric dipole moment (dipole operator $\propto\tilde{\tau}_z^{(i)}$) resulting in an electric dipole interaction between the DQD charge degree of freedom and the resonator photons \cite{scarlino2021,burkard2020},
\begin{equation}
\tilde{H}_{\mathrm{int}}^{(i)} = g_c(a + a^\dagger)\tilde{\tau}_z^{(i)}.
\end{equation} 
In recent experiments a coupling strength of $g_c/2\pi \approx \SI{513}{\mega\hertz}$ for a hole-charge qubit was reported \cite{yu2023}.

\section{\label{sec:idealconsiderations}Entangling two-qubit gates}

In this section, we derive an effective Hamiltonian describing the resonator-mediated interaction between two charge qubits and assess the potential of this interaction for quantum information applications. 
As a first step, the states constituting the individual charge qubits are set as the two eigenstates of $\tilde{H}_{\mathrm{DQD}}^{(i)}$,
\begin{align}
    \ket{+^{(i)}} & = \cos\tfrac{\theta^{(i)}}{2} \ket{L^{(i)}} + \sin\tfrac{\theta^{(i)}}{2} \ket{R^{(i)}},     \\
    \ket{-^{(i)}} & = \sin\tfrac{\theta^{(i)}}{2} \ket{L^{(i)}} - \cos\tfrac{\theta^{(i)}}{2} \ket{R^{(i)}},
\end{align}
with the orbital angle $\theta^{(i)}$ defined by ${\tan\theta^{(i)} = 2 t_c^{(i)} / \varepsilon^{(i)}}$. The corresponding eigenvalues are 
\begin{equation}
    \pm\Omega^{(i)}/2= \pm \frac{1}{2}\sqrt{(\varepsilon^{(i)})^2 +  (2t_c^{(i)})^2},
    \label{eq:Omega}
\end{equation}
and in this basis the interaction between resonator and qubit $\tilde{H}_{\mathrm{int}}^{(i)}$ reads
\begin{align}
    H^{(i)}_{\mathrm{int}} = g_c (a + a^\dagger)( \sin\theta^{(i)} \tau_x^{(i)} + \cos\theta^{(i)} \tau_z^{(i)} ),
    \label{eq:Hint}
\end{align}
where the Pauli matrices $\tau_j^{(i)}$ operate on the qubit states as $\tau_z^{(i)} \ket{\pm^{(i)}} = \pm \ket{\pm^{(i)}}$.
Eq.~\eqref{eq:Hint} clearly shows that the transverse qubit-resonator interaction, represented by the term $\propto \tau_x^{(i)}$, can be switched on and off by modulating the DQD detuning $\varepsilon^{(i)}$ and the tunnel coupling strength $t_c^{(i)}$. This allows individual qubits to be decoupled from the resonator and placed into an idling mode.

Readout of an individual charge qubit can be achieved by decoupling it from the resonator and adiabatically modulating the DQD detuning towards a strong bias, resulting in a mapping of the state $\ket{+^{(i)}}$ ($\ket{-^{(i)}}$) to the state $\ket{L^{(i)}}$ or $\ket{R^{(i)}}$ ($\ket{R^{(i)}}$ or $\ket{L^{(i)}}$), depending on the specific form of the detuning pulse. The states $\ket{L^{(i)}}$ and $\ket{R^{(i)}}$ can then be differentiated by measuring the charge occupation of the individual quantum dots using a quantum point contact (QPC) \cite{dicarlo2004}, a single electron transistor (SET) \cite{Lu2003}, or radiofrequency gate sensors \cite{colless2013}.

Alternatively, in a setup where the charge qubit of interest is dispersively coupled to the resonator while the second qubit is decoupled from the resonator, the DQD system can be measured directly in the basis $\{\ket{+^{(i)}}, \ket{-^{(i)}}\}$ by assessing the qubit state-dependent resonator frequency shift \cite{scarlino2019}.

We note that the energy levels of symmetric DQDs ($\varepsilon^{(i)}=0$) are to first order insensitive to fluctuations in the detuning $\varepsilon^{(i)}$, i.e., $\left.\left(\partial \Omega^{(i)}/\partial \varepsilon^{(i)}\right)\right|_{\varepsilon^{(i)}=0}=0$. Thus, adjusting a DQD to $\varepsilon^{(i)}=0$ means tuning the qubit to a charge-noise sweet spot \cite{benito2017}. 
In the following we consider identical qubits at zero detuning, i.e., $\varepsilon^{(i)}=\varepsilon=0$ and $t_c^{(i)}=t_c$, unless noted otherwise. 
Then the interaction Hamiltonian $\tilde{H}_{\mathrm{int}}^{(i)}$ \eqref{eq:Hint} simplifies to
\begin{equation}
H_{\mathrm{int}}^{(i)} = g_c \left(a + a^\dagger \right)\tau^{(i)}_x.
\label{eq:Hintplusminus}
\end{equation}
 The basis change leaves $H_c$ unaffected. Eq.~\eqref{eq:Hintplusminus} already suggests that there is a second-order process coupling the states $|+,-\rangle$ and $|-,+\rangle$, where $|i,j\rangle$ indicates the  state of charge qubit 1 and 2, respectively, as given by $i$ and $j$.

The charge qubits are resonant with the resonator mode if $\Omega=2t_c=\omega_c$. Here, we focus on the dispersive regime $|2t_c - \omega_c| \gg g_c$ and low temperatures such that the resonator is empty. In this regime we can derive an effective Hamiltonian by applying a Schrieffer-Wolff-Transformation (see Appendix \ref{sec:swtrafo}) to decouple the subspace with zero photons from the one with higher photon numbers. For the zero photon subspace, we find the effective two-qubit Hamiltonian,
\begin{equation}
    \bar{H}_d = \sum_{i=1}^2 \frac{E}{2} \tau_{z}^{(i)} + g_{\mathrm{eff}} \tau_{x}^{(1)}\tau_{x}^{(2)},\label{eq:hameff}
\end{equation}
with  
\begin{equation}
    E = 2t_c - \frac{g_c^2}{\omega_c-2t_c}+ \frac{g_c^2}{\omega_c+2t_c},
\end{equation}
and the effective qubit-qubit coupling strength
\begin{equation}
    g_{\mathrm{eff}} = - \frac{g_c^2}{\omega_c-2 t_c} - \frac{g_c^2}{\omega_c+2 t_c}. \label{eq:geff}
\end{equation}
The dynamics generated by the diagonal terms of \eqref{eq:hameff} can be captured by a transformation into the interaction picture with respect to $\sum_{i=1}^2 E  \tau_{z}^{(i)}/2$. In this frame \eqref{eq:hameff} reads
\begin{align}
    H &= g_{\mathrm{eff}}(\tau_{+}^{(1)}\tau_{-}^{(2)} + \tau_{-}^{(1)}\tau_{+}^{(2)}) 
    \nonumber \\
    &+ g_{\mathrm{eff}}(e^{2iEt}\tau_{+}^{(1)}\tau_{+}^{(2)} + e^{-2iEt}\tau_{-}^{(1)}\tau_{-}^{(2)}).
   \label{eq:hideal}
\end{align}
The effective interaction between the two charge qubits described by the first term in Eq.~\eqref{eq:hideal} leads to two-qubit dynamics characterized by a time scale $\propto 1/g_{\mathrm{eff}}$. In the dispersive regime one has $|2E|\approx |2(2t_c)|>|2t_c-\omega_c|\gg g_c$ and therefore $g_{\mathrm{eff}}\ll g_c$ according to Eq.~\eqref{eq:geff} as well as $g_{\mathrm{eff}}\ll |2E|$ such that applying the rotating wave approximation (RWA) is justified. 
With the RWA applied the second term is neglected and it is straightforward to calculate the time evolution operator generated by the Hamiltonian \eqref{eq:hideal},
\begin{equation}
    U(t,0) = \begin{pmatrix*}
        1 & 0 & 0 & 0 \\
        0 & \cos(g_{\mathrm{eff}} t) & -i\sin(g_{\mathrm{eff}} t) & 0 \\
        0 & -i\sin(g_{\mathrm{eff}} t) & \cos(g_{\mathrm{eff}} t) & 0 \\
        0 & 0 & 0 & 1 \\
    \end{pmatrix*},\label{eq:u}
\end{equation}
with respect to the basis $\{\ket{+,+},\ket{+,-},\ket{-,+},\ket{-,-}\}$.
The time evolution for the particular evolution times 
\begin{equation}
  t_{i\mathrm{SWAP}}=\pi / 2 |g_{\mathrm{eff}}|,
  \label{eq:tiSWAP}
\end{equation} 
and 
\begin{equation}
    t_{\sqrt{i\mathrm{SWAP}}}=\pi / 4|g_{\mathrm{eff}}|,
     \label{eq:tsqrtiSWAP}
\end{equation}
results in 
\begin{equation}
U\left(t_{i\mathrm{SWAP}},0\right) =\begin{pmatrix}
       1 & 0 &0 &0 \\
       0 & 0&\pm i  &0 \\
       0& \pm i& 0 &0 \\
       0 & 0 &0 &1
    \end{pmatrix},
    \label{eq:UiSWAP}
\end{equation} and 
\begin{equation}
  U\left(t_{\sqrt{i\mathrm{SWAP}}},0\right)=  \begin{pmatrix}
		1 & 0 & 0 & 0 \\
		0 & 1/\sqrt{2} & \pm i /\sqrt{2}  & 0  \\
		0&  \pm i /\sqrt{2}   & 1/\sqrt{2} & 0 \\
		0 & 0 & 0 & 1 
	\end{pmatrix},
  \label{eq:UsqrtiSWAP}
\end{equation} 
respectively, 
where the the sign $-(+)$ is obtained for $g_{\mathrm{eff}}/|g_{\mathrm{eff}}|=+(-)1$. 

The Makhlin invariants \cite{Makhlin2002} of \eqref{eq:UiSWAP}
and \eqref{eq:UsqrtiSWAP} are independent of the sign of the off-diagonal elements, and therefore the respective operations with opposite sign are similar up to single qubit rotations. Thus, we refer to  \eqref{eq:UiSWAP} as an $i\mathrm{SWAP}$ gate and to \eqref{eq:UsqrtiSWAP} as an $\sqrt{i\mathrm{SWAP}}$ quantum gate.

\section{\label{sec:errors}Noise analysis}
The qubits and the microwave resonator couple to their respective environment. Therefore, it is necessary to analyze the impact of processes resulting from these couplings on the gate performance in detail. Here, we provide a detailed discussion of the two dominant noise processes: resonator photon decays and charge qubit dephasing. The effect of both noise processes on the gate performance is analyzed by calculating the average gate fidelity $\bar{F}$ introduced in Ref.~\cite{nielsen2002} and briefly summarized in Appendix~\ref{sec:fidelity}.

Charge qubit relaxation does not significantly impact gate fidelity. This is evindenced by reports of DQD charge qubit relaxation times \( T_1 = 45 \,\mu\mathrm{s} \) \cite{wang2013}, while in our study, as detailed below, gate times range from a few nanoseconds to approximately 125 ns (see Fig.~\ref{fig:dephasingcom}).

\subsection{Resonator damping}

\begin{figure}[b]
    \centering
    \includegraphics[width=\columnwidth]{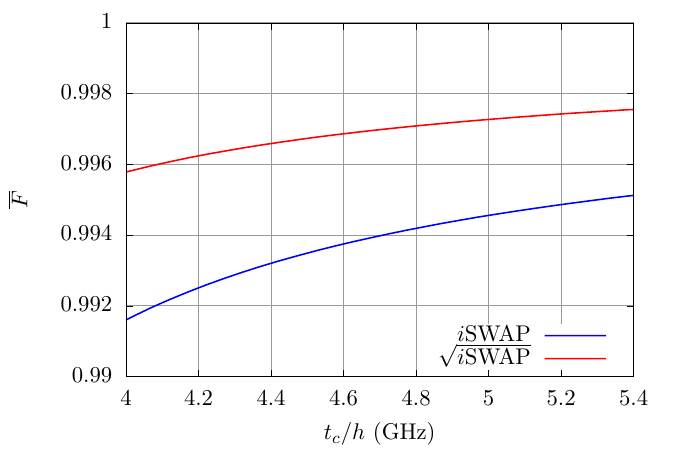}
    \caption{Average gate fidelity of the $i\mathrm{SWAP}$ (blue) and $\sqrt{i\text{SWAP}}$ (red) quantum gates accounting for resonator damping as a function of the tunnel coupling strength $t_c$. Here $g_c/ (2\pi) = \SI{513}{\mega\hertz}$, $\omega_c / (2\pi) = \SI{5.428}{\giga\hertz}$ and $\kappa  / (2\pi)= \SI{14}{\mega\hertz}$ are chosen \cite{yu2023}.\label{fig:lfc}}
\end{figure}

The interaction of the resonator with its environment is described by a coupling between the resonator mode and an electromagnetic bath. While, in practice, experiments are typically performed at cryogenic temperatures $T\leq10\,\mathrm{mK}$ \cite{mi2017}, for simplicity, we assume that the environment is held at zero temperature. In this limit the environment resides in its ground state and the conservation of energy prohibits processes increasing the energy of the DQD-resonator system due to interactions with the environment. Moreover, there is no thermal population of the resonator, i.e., $\langle a^{\dagger}a\rangle=0$.
The reasoning we present in Appendix~\ref{sec:appcavitydamping} demonstrates that the dissipative dynamics due to resonator damping are described by the Lindblad master equation 
\begin{equation}
    \dot{\rho} = -i[H, \rho] + \frac{\kappa g_c^2}{(\omega_c - 2t_c)^2} \mathcal{D}[\tau_-^{(1)} + \tau_-^{(2)}](\rho) \,, \label{eq:dglcavity}
\end{equation}
with the density matrix $\rho$, the dissipator superoperator $\mathcal{D}[c](\rho) = c\rho c^\dagger - \{c^\dagger c, \rho\}/2$ and the resonator photon loss rate $\kappa$. Using \eqref{eq:dglcavity}, the average fidelity of the $i$SWAP and  $\sqrt{i\text{SWAP}}$ gates as a function of the gate time is given by
\begin{equation}
    \bar{F} = \frac{1}{10}\left[2+(1+x)(1+2xy+y^2)\right], \label{eq:fcavity}
\end{equation}
with $y = \exp\{-t \frac{\kappa g_c^2}{(\omega_c - 2t_c)^2}\}$ and $x = \sin(|g_{\mathrm{eff}}| t)$ for the $i$SWAP gate and $x = \sin(|g_{\mathrm{eff}}| t + \pi/4)$ for the $\sqrt{i\text{SWAP}}$ gate. The average gate fidelity for both gates is presented in Fig.~\ref{fig:lfc} as a function of the tunnel coupling strength $t_c$ with which the resonator-DQD detuning is controlled. If the system is operated deep in the dispersive regime characterized by $|2t_c-\omega_c|\gg g_c$, e.g. as in Ref.~\cite{yu2023}, the decay rate in \eqref{eq:dglcavity} can be as small as $20\,\mathrm{kHz}$ resulting in an average gate fidelity exceeding $99\%$.
Resonator losses have the largest impact on the average gate fidelity at the resonance $2t_c = \omega_c$. This is not surprising because there is strong charge-photon hybridization at the resonance and, therefore, the gate is more susceptible to photon losses. The higher gate fidelity of the $\sqrt{i\mathrm{SWAP}}$ gate compared to the $i\mathrm{SWAP}$ gate can be attributed to the difference in the gate times, i.e., $2\, t_{\sqrt{i\mathrm{SWAP}}}=t_{i\mathrm{SWAP}}$.

\subsection{Charge qubit dephasing \label{sec:charge_qubit_dephasing}}

Qubit dephasing tends to be the limiting factor for the coherence of charge qubits. 
According to the detailed derivation presented in App.~\ref{sec:appendixdephasing}, charge qubit dephasing is described by the master equation
\begin{align}
    \dot{\rho} &= -i[H,\rho] + \frac{\gamma_{\phi}(t_c)}{2}  \sum_{i=1}^2 \mathcal{D}[\left(1 + \alpha_1\right) \tau_z^{(i)} \nonumber \\
    &+ \alpha_2 (1 + \tau_{+}^{(1)}\tau_{-}^{(2)} + \tau_{-}^{(1)}\tau_{+}^{(2)})](\rho), \label{eq:dglphonon}
\end{align}
where the explicit definitions of the coefficients $\alpha_1$ and $\alpha_2$ in terms of the system parameters are listed in Appendix~\ref{sec:appendixdephasing}.
As explained in detail in App.~\ref{app:charge_dephasing}, the charge dephasing rate $\gamma_{\phi}(t_c)$ at the charge-noise sweet spot ($\varepsilon=0$) is inversely proportional to the tunnel coupling strength,  $\gamma_{\phi}\propto 1/t_c$, with a reported value of $\gamma_{\phi} / (2\pi) = \SI{9.9}{\mega\hertz}$ for $2t_c / h = \SI{19.2}{\giga\hertz}$ \cite{yu2023}. When writing the explicit form of the dissipator in \eqref{eq:dglphonon}, one finds terms $\mathcal{O}(\alpha_i^0)$, terms $\propto \alpha_i\propto g_c^2$, and terms $\propto \alpha_i\alpha_j\propto g_c^4$ with $i,j\in\{1,2\}$. In line with \eqref{eq:dglcavity} we consider the terms $\mathcal{O}(g_c^2)$, while the terms $\mathcal{O}(g_c^4)$ are neglected in the following. We point out that keeping the $\mathcal{O}(g_c^4)$ terms would require including higher order corrections in the transformation~\eqref{eq:tauzSWexpansion} because these corrections can give rise to additional terms $\mathcal{O}(g_c^4)$ in \eqref{eq:dglphonon}.

As the next step, \eqref{eq:dglphonon} is employed to calculate the average gate fidelity for the $i\mathrm{SWAP}$ and the $\sqrt{i\mathrm{SWAP}}$ gate,
\begin{align}
    \bar{F}_{i\mathrm{SWAP}} =& \frac{1}{20} \left[ 7 + 3y_\gamma^2 - 2 y_\gamma \cos(\beta t) + 8 x y_\gamma z\right], \label{eq:fdephasing}\\
    \bar{F}_{\sqrt{i\mathrm{SWAP}}} =& \frac{1}{20} \Big[ 7 + 3y_\gamma^2 
    +  4 y_\gamma |g_{\mathrm{eff}}| \frac{\sin(\beta t)}{\beta} + 8 x y_\gamma z \Big],
\end{align}
with ${y_\gamma = \exp \{ - \gamma_{\phi}(t_c) ( 1 + 2\alpha_1) t\}}$,
${\beta = \sqrt{4 g_{\mathrm{eff}}^2-\gamma_{\phi}^2(t_c) (1+2\alpha_1)^2}}$,
and ${z = \cosh(\gamma_{\phi}(t_c) \alpha_2 t)}$. 

\begin{figure}
    \centering
    \includegraphics[width=\columnwidth]{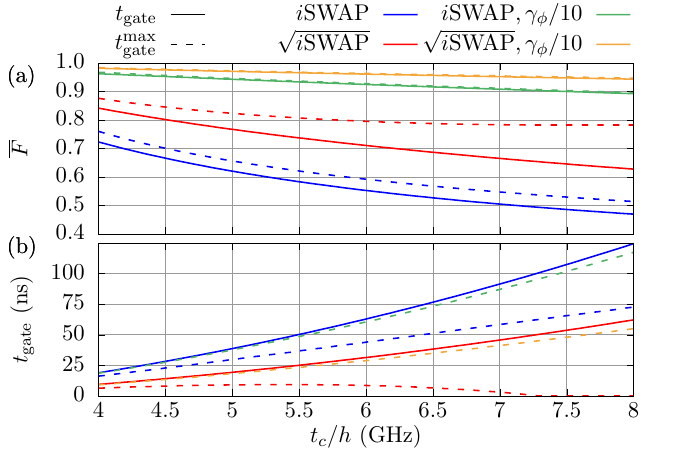}
    \caption{(a) Average gate fidelity of the $i\mathrm{SWAP}$ and the $\sqrt{i\mathrm{SWAP}}$ quantum gate affected by charge dephasing and (b) the corresponding gate time as a function of the tunnel coupling strength $t_c$. For the solid lines the gate time is chosen as in the Eqs.~\eqref{eq:tiSWAP} and \eqref{eq:tsqrtiSWAP}, while the dashed lines correspond to $t_{i\mathrm{SWAP}}^{\mathrm{max}}$ and $t_{\sqrt{i\mathrm{SWAP}}}^{\mathrm{max}}$, i.e., the numerically obtained evolution times for which the average gate fidelity $\bar{F}$ is maximized. The gate times $t_{i\mathrm{SWAP}}$ (Eq.~\eqref{eq:tiSWAP}) and $t_{\sqrt{i\mathrm{SWAP}}}$ (Eq.~\eqref{eq:tsqrtiSWAP}) are independent of the dephasing rate. Thus, the solid blue and green lines, and the solid red and yellow lines in (b) overlap. The charge dephasing rate given in \eqref{eq:gammaphioftc} underlying simulations for the red and blue curves is in line with recent experiments \cite{yu2023}. The remaining parameters are chosen as in Fig.~\ref{fig:lfc}.}
    \label{fig:dephasingcom}
\end{figure}

Figure~\ref{fig:dephasingcom}(a) shows the average fidelity of the $i\mathrm{SWAP}$ and  $\sqrt{i\mathrm{SWAP}}$ gates as a function of the tunnel coupling strength $t_c$. The blue (red) line gives the average gate fidelity for the evolution time set to $t_{i\mathrm{SWAP}} \eqref{eq:tiSWAP}$ ($t_{\sqrt{i\mathrm{SWAP}}}$ \eqref{eq:tsqrtiSWAP}) for which the respective gate is realized in the absence of decoherence processes according to \eqref{eq:UiSWAP}(\eqref{eq:UsqrtiSWAP}). The dashed lines, however, show the maximum gate fidelity for the specific set of system parameters obtained by maximizing the gate fidelity as a function of the evolution time. The green (yellow) line was obtained by assuming a charge dephasing rate an order of magnitude smaller than the values achieved in recent experiments. A comparison between the evolution time for which maximal gate fidelity $t_{i\mathrm{SWAP}}^{\mathrm{max}}$ ($t_{\sqrt{i\mathrm{SWAP}}}^{\mathrm{max}}$) is achieved and $t_{i\mathrm{SWAP}}$($t_{\sqrt{i\mathrm{SWAP}}}$) is presented in Fig.~\ref{fig:dephasingcom}(b).

Figure~\ref{fig:dephasingcom}(a) unveils that charge dephasing reduces the gate fidelity in the dispersive regime significantly. Due to the scaling $\gamma_{\phi}(t_c)\propto 1/t_c$, the dephasing rate increases with a decreasing tunnel coupling strength. However, the Figure shows a reduction of the average gate fidelity with rising tunnel coupling strength. A closer look at the gate times as a function of the tunnel coupling strength presented in  Fig.~\ref{fig:dephasingcom}(b) provides an explanation for this observation: Increasing the tunnel coupling strength results in a longer gate time such that the system experiences charge dephasing at a reduced rate for a longer time period, whereby the longer gate time outweighs the benefits of a reduced dephasing rate. 
Moreover, Fig.~\ref{fig:dephasingcom}(b) demonstrates that the maximal gate fidelity is achieved for evolution times shorter than the gate time in the ideal decoherence free scenario, $t_{i\mathrm{SWAP}}$($t_{\sqrt{i\mathrm{SWAP}}}$).

Analyzing Fig.~\ref{fig:dephasingcom} for a specific value of the tunnel coupling strength $t_c$ unveils a higher average fidelity of the $\sqrt{i\mathrm{SWAP}}$ gate than the $i\mathrm{SWAP}$ gate. This property can again be attributed to the difference in gate time. 

Inspecting the red dashed lines in Fig.~\ref{fig:dephasingcom}, one observes an initially surprising behavior. In the highly dispersive regime with $t_c/h \gtrsim \SI{7.25}{\giga\hertz}$, the average gate fidelity approaches a steady value while the gate time goes to zero. A vanishing gate time implies that the initial state $\ket{\psi_0}$ is not altered in any way, i.e., the map $\mathcal{E}$ introduced in Appendix~\ref{sec:fidelity} acts on the initial state as $\mathcal{E}(\ket{\psi}\bra{\psi})=\ket{\psi}\bra{\psi}$. In this case, according to \eqref{eq:averagefidelity}, the average gate fidelity for the $\sqrt{i\mathrm{SWAP}}$ gate amounts to 
\begin{align}
    \bar{F}(\mathcal{E}, U(t_{\sqrt{i\mathrm{SWAP}}},0)) = \frac12 + \frac{\sqrt{2}}5\approx 0.78.
\end{align}
This result agrees with the value to which the red dashed line in Fig.~\ref{fig:dephasingcom}(a) converges.  
The above reasoning allows us to conclude that an average gate fidelity of the $\sqrt{i\mathrm{SWAP}}$ gate exceeding the ``initial value" of 0.78 for $t>0$ requires $(\mathrm{d}\bar{F}_{\sqrt{i\mathrm{SWAP}}}/\mathrm{d}t)|_{t=0} \neq 0$. This condition is fulfilled for  
\begin{align}
|g_{\mathrm{eff}}|>\frac{\left(1+\sqrt{2}\right)\left(1+2\alpha_1\right)}{2}\gamma_{\phi}(t_c).
\end{align}
 The comparison of the Figs.~\ref{fig:lfc} and \ref{fig:dephasingcom} shows that, in the dispersive regime, charge dephasing is the dominant decoherence mechanism. 

At first sight, our results suggest an increased fidelity in the case that the coupling between the qubits and the resonator is closer to the resonance. On resonance, however, the realization of the $i\mathrm{SWAP}$ or the $\sqrt{i\mathrm{SWAP}}$ gate by simultaneously coupling two identical charge qubits to a resonator is not possible as explained in the following. In the resonant regime, the system is described by the Hamiltonian 
\begin{align}
    H_{\mathrm{res}}=g_c \sum_{i=1}^2\left(\tau_+^{(i)}a+\tau_-^{(i)}a^{\dagger}\right),
\end{align}
in a rotating reference frame with respect to $\sum_{i=1}^2 t_c\tau_z^{(i)}$  and with the RWA applied. This Hamiltonian generates the time evolution
\begin{align}
    U_{\mathrm{res}}(t)=\exp(-i H_{\mathrm{res}} t),
\end{align}
which reduces to 
\begin{align}
  &U^{0p}_{\mathrm{res}}(t)=P_{0}U_{\mathrm{res}}(t)P_{0}=\nonumber\\  
  &\begin{pmatrix}
        \frac{1}{3}\left[2+\cos\left(\sqrt{6} g_c t\right)\right] & 0&0 &0 \\
        0 & \cos^2\left(\frac{g_c}{\sqrt{2}}t\right) &-\sin^2\left(\frac{g_c}{\sqrt{2}}t\right) & 0 \\
         0 &-\sin^2\left(\frac{g_c}{\sqrt{2}}t\right) & \cos^2\left(\frac{g_c}{\sqrt{2}}t\right)& 0 \\
         0 & 0& 0 & 1
    \end{pmatrix},
\end{align}
within the zero-photon subspace corresponding to the projection operator $P_{0}=\ket{0}\bra{0}$.
The generally non-vanishing matrix elements connecting the zero-photon subspace to states with higher photon numbers are
\begin{align}
\bra{-,-,1}U_{\mathrm{res}}(t)\ket{-,+,0}&=\bra{-,-,1}U_{\mathrm{res}}(t)\ket{+,-,0}\nonumber\\
    &=-i\sin\left(\sqrt{2}g_c t\right)/\sqrt{2},
     \label{eq:matrixelementleakage1}\\
    \bra{-,+,1}U_{\mathrm{res}}(t)\ket{+,+,0}&=\bra{+,-,1}U_{\mathrm{res}}(t)\ket{+,+,0}\nonumber\\
    &=-i\sin\left(\sqrt{6}g_ct\right)/\sqrt{6},
    \label{eq:matrixelementleakage2}\\
    \bra{-,-,2}U_{\mathrm{res}}(t)\ket{+,+,0}&=-\frac{2\sqrt{2}}{3}\sin^2\left(\frac{\sqrt{3}g_c}{\sqrt{2}}t\right).
     \label{eq:matrixelementleakage3}
\end{align}
A closer look unveils that there is no evolution time except $t=0$ for which the matrix elements \eqref{eq:matrixelementleakage1}-\eqref{eq:matrixelementleakage3} vanish simultaneously, indicating an unavoidable leakage from the zero photon subspace to states with higher photon numbers, and  thus implementing a quantum gate restricted to the zero photon subspace is not feasible in the resonant regime. 

\section{\label{sec:conclusion} Conclusion}

We demonstrated that a microwave resonator mediates a long-range interaction between two charge qubits defined in two spatially separated Si DQDs, each hosting a single charge carrier, either an electron or a hole, while being dispersively coupled to the same resonator mode.
We found that this interaction can be harnessed to implement both the $i\mathrm{SWAP}$ and the $\sqrt{i\mathrm{SWAP}}$ two-qubit quantum gate within a gate time in the range of a few nanoseconds.

A detailed discussion of the most prominent noise sources, i.e., resonator photon losses and charge qubit dephasing, have shown that in the dispersive regime, charge dephasing is the dominant decoherence mechanism. Assuming a state-of-the art resonator decay rate and charge qubit dephasing rate, the predicted average gate fidelities are below 90\%. However, a reduction of the charge qubit dephasing rate by one order of magnitude is predicted to result in gate fidelities surpassing 95\%. 
Such a reduction is a plausible scenario in the near future, as understanding the origin of charge noise—the primary cause of charge and spin dephasing—is an active field of research \cite{connors2022,ye2024a}. This ongoing research has already led to promising approaches to mitigate charge noise \cite{paqueletwuetz2023,elsayed2024,ye2024}.

 \begin{acknowledgments}
 J.M. and G.B. acknowledge the support by ARO grant number W911NF-15-1-0149.
  \end{acknowledgments}

\appendix

\section{\label{sec:swtrafo}Schrieffer-Wolff-Transformation to eliminate excited resonator states}

The dynamics of the system consisting of two charge qubits coupled to a single mode of a microwave resonator is captured by the Hamiltonian
${H=H_c+\sum_{i=1}^2H_{\mathrm{DQD}}^{(i)}+H_{\mathrm{int}}^{(i)}}$, with ${H_{\mathrm{DQD}}^{(i)}=\frac{\Omega^{(i)}}{2}\tau_{z}^{(i)}}$ and ${H_{\mathrm{int}^{(i)}} = g_c (a+a^\dagger)(\sin \theta^{(i)} \tau_x^{(i)} + \cos\theta^{(i)} \tau_z^{(i)})}$. In order to decouple the dynamics of the zero-photon subspace from the excited resonator states, we first divide $H$ into a diagonal part $H_0=H_c+\sum_{i=1}^2H_{\mathrm{DQD}}^{(i)}$, and an off-diagonal part ${V = V_{\mathrm{d}} + V_{\mathrm{od}}}$ \cite{bravyi2011}. Using the projectors $P_0=\ket{0}\bra{0}$ and $Q_0 = 1 - P_0$, the block diagonal and block off-diagonal parts of $V$ can be obtained as
\begin{align}
    V_{\mathrm{d}} &= P_0 V P_0 + Q_0 V Q_0, \\
    V_{\mathrm{od}} &= P_0 V Q_0 + Q_0 V P_0.
\end{align}
Next, one can apply a Schrieffer-Wolff transformation with antihermitian generator $S$ to find the effective Hamiltonian 
\begin{align}
    \bar{H} = e^{S} H e^{-S}.
\end{align}
Here, we follow the perturbative approach ${S = S_1 + S_2 + \dots}$, with $S_n\sim V^n$ discussed in Ref.~\cite{bravyi2011}, allowing for the decoupling of the subspaces with different photon numbers up to a desired order in $V$. 
The first two contributions to the generator $S$, $S_1$ and $S_2$, must obey the relations \cite{bravyi2011},
\begin{align}
    [H_0, S_1] & = V_{\mathrm{od}},     \label{eq:defS1}  \\
    [H_0, S_2] & = - [V_{\mathrm{d}},S_1]. \label{eq:defS2}
\end{align}
Then, the effective Hamiltonian in the invariant subspace of $P_0$, up to second order in the perturbation $V$, reads
\begin{align}
    \bar{H}_d= H_0 P_0 + P_0VP_0 + \frac12 P_0[S_1,V_{\mathrm{od}}]P_0.
    \label{eq:HeffSW}
\end{align}
In general, an operator $X$ transforms under the transformation generated by $S$ as 
\begin{align}
    e^{S}Xe^{-S}&=\sum_{j=0}^{\infty}\frac{1}{j!}[X,-S]^{(j)}\nonumber\\
    &\approx X-[X,S_1]-[X,S_2]+\frac{1}{2}[X,S_1]^{(2)}, 
    \label{eq:SWoperators}
\end{align}
where the second line describes the contributions up to second order in the perturbation $V$ and ${[A,B]^{(j)}=[...[[A\underbrace{,B],B],...B]}_{j\mathrm{- times}}}$. 

In the general case of distinct qubits, i.e., $\varepsilon^{(1)}\neq \varepsilon^{(2)}$ and  $t_c^{(1)}\neq t_c^{(2)}$, the Eqs.~\eqref{eq:defS1} and \eqref{eq:defS2} are solved by  
\begin{align}
    S_1 &=  \sum_{i=1}^2 \frac{g_c\sin\theta^{(i)}}{\omega_c + \Omega^{(i)}} \left(\ket{1}\bra{0} \tau_+^{(i)} - \ket{0}\bra{1}\tau_-^{(i)}\right) \nonumber \\
    &+ \sum_{i=1}^2 \frac{g_c\sin\theta^{(i)}}{\omega_c -\Omega^{(i)}} \left(\ket{1}\bra{0} \tau_-^{(i)} - \ket{0}\bra{1}\tau_+^{(i)}\right) \nonumber \\
    &+ \sum_{i=1}^2 \frac{g_c\cos\theta^{(i)}}{\omega_c}\left(\ket{1}\bra{0}  - \ket{0}\bra{1}\right)\tau_z^{(i)}, \label{eq:s1erstetrafo}
\end{align}
and 
\allowdisplaybreaks{
\begin{align}
S_2=\sqrt{2}\ket{0}&\bra{2} \Bigg(\sum_{i=1}^2\frac{g_c^2\cos^2\theta^{(i)}}{2\omega_c^2} \nonumber\\
&+\tau_z\sigma_z\frac{g_c^2\cos\theta^{(1)}\cos\theta^{(2)}}{\omega_c^2} \nonumber\\
&+\sum_{i=1}^2\frac{1+\tau_z^{(i)}}2\frac{g_c^2\sin^2\theta^{(i)}}{2\omega_c(\omega_c-\Omega^{(i)})} \nonumber\\
&+\sum_{i=1}^2\frac{1-\tau_z^{(i)}}2\frac{g_c^2\sin^2\theta^{(i)}}{2\omega_c(\omega_c+\Omega^{(i)})} \nonumber\\
&-\sum_{i=1}^2\tau_-^{(i)}\frac{g_c^2\Omega^{(i)}\cos\theta^{(i)}\sin\theta^{(i)}}{\omega_c(2\omega_c+\Omega^{(i)})(\omega_c+\Omega^{(i)})} \nonumber\\
&-\sum_{i=1}^2\tau_+^{(i)}\frac{g_c^2\Omega^{(i)}\cos\theta^{(i)}\sin\theta^{(i)}}{\omega_c(2\omega_c-\Omega^{(i)})(\omega_c-\Omega^{(i)})} \nonumber\\
&+\sum_{i=1}^2\tau_-^{(i)}\tau_-^{(j\neq i)}\frac{g_c^2\sin\theta^{(i)}\sin\theta^{(j\neq i)}}{(\omega_c+\Omega^{(i)})(\omega_c+\Omega^{(j\neq i)})} \nonumber\\
&+\sum_{i=1}^2\tau_-^{(i)}\tau_+^{(j\neq i)}\frac{g_c^2\sin\theta^{(i)}\sin\theta^{(j\neq i)}}{(\omega_c+\Omega^{(i)})(\omega_c-\Omega^{(j\neq i)})} \nonumber\\
&+\sum_{i=1}^2\tau_+^{(i)}\tau_z^{(j\neq i)}\frac{g_c^2\sin\theta^{(i)}\cos\theta^{(j\neq i)}}{\omega_c(\omega_c-\Omega^{(i)})} \nonumber\\
&+\sum_{i=1}^2\tau_-^{(i)}\tau_z^{(j\neq i)}\frac{g_c^2\sin\theta^{(i)}\cos\theta^{(j\neq i)}}{\omega_c(\omega_c+\Omega^{(i)})} \Bigg),\nonumber\\
&- \mathrm{h.c},
\end{align}}
respectively.

 Given the above expression for $S_1$, it is straightforward to determine the explicit form of the effective Hamiltonian $\bar{H}_d$~\eqref{eq:HeffSW} up to second order in the perturbation $V$,
\begin{align}
    \bar{H}_d &= \sum_{i=1}^2 \frac{E^{(i)}}{2} \tau_{z}^{(i)} + g_{\mathrm{eff}} \tau_{x}^{(1)}\tau_{x}^{(2)} + g_1 \tau_z^{(1)}\tau_z^{(2)} \nonumber \\ 
    &+ \sum_{i=1}^2 g_2^{(i)} \tau_x^{(i)} + \sum_{i=1}^2 g_3^{(i)} \tau_z^{(i)} \tau_x^{(j\neq i)},
    \label{eq:heff}
\end{align}
up to an unimportant constant term that is neglected here. The coefficients are given by
\allowdisplaybreaks{
\begin{align}
    g_{\mathrm{eff}}       & = g_c^2\sin\theta^{(1)}\sin\theta^{(2)} \sum_{i=1}^2 \frac{\omega_c}{\omega_c^2 - (\Omega^{(i)})^2}, \label{eq:geffapp} \\
    E^{(i)} & = \Omega^{(i)} - \frac{g_c^2\sin^2\theta^{(i)}}{\omega_c-\Omega^{(i)}} + \frac{g_c^2\sin^2\theta^{(i)}}{\omega_c+\Omega^{(i)}}, \\
    g_1 &= - \frac{2g_c^2 \cos\theta^{(1)}\cos\theta^{(2)}}{\omega_c}, 
    \label{eq:g_1}\\
    g_2^{(i)} &= \frac{g_c^2 \Omega^{(i)} \cos\theta^{(i)}\sin\theta^{(i)}}{\omega_c^2 - (\Omega^{(i)})^2}, \\
    g_3^{(i)} &= g_c^2 \cos\theta^{(i)}\sin\theta^{(j\neq i)} \left(\frac{1}{\omega_c} + \frac{\omega_c}{\omega_c^2 - (\Omega^{(i)})^2}\right).
\end{align}}
To arrive at the equations \eqref{eq:hameff}--\eqref{eq:geff} in the main text, identical qubits operated at the charge noise sweet spot $\varepsilon^{(i)} = 0$ are considered. In particular, this implies $\cos\theta^{(i)} = 0$ and $\sin\theta^{(i)} = 1$ as well as $\Omega^{(i)}=2t_c$ for $i=1,2$. We note that the presence of the ZZ interaction in equation~\eqref{eq:heff} is unfavorable for the implementation of the two-qubit gates discussed in the main text (see also \cite{Moskalenko2022}). However, the explicit expression for the associated coupling strength, $g_1$ \eqref{eq:g_1}, provides a solution to this issue. When operating the qubits at the charge noise sweet spot with $\varepsilon^{(i)} = 0$, one finds that $g_1$ vanishes. Additionally, near the charge noise sweet spot, the ZZ interaction is strongly suppressed as $g_1 \propto \varepsilon^{(1)}\varepsilon^{(2)}$. Furthermore, operating at the charge noise sweet spot maximizes the desired effective coupling $g_{\mathrm{eff}}$ between the two qubits.

We note that the perturbative approach underlying the derivation of \eqref{eq:heff} requires $\left|g_c \sin\theta^{(i)}\right|\ll \left|\Omega^{(i)}-\omega_c\right|,\left|\Omega^{(i)}+\omega_c\right|$. Hence, $\bar{H}_d$ is not valid in parameter domains with $\Omega^{(i)}\approx \omega_c$ where the qubits are close to resonance with the resonator mode.

\section{\label{sec:fidelity}Fidelity}

Given the desired time evolution $\ket{\psi}\bra{\psi} \to U\ket{\psi}\bra{\psi}U^{\dagger}$ of a quantum system and the real implementation described by the map $\ket{\psi}\bra{\psi} \to \mathcal{E}(\ket{\psi}\bra{\psi})$, we can measure the deviation of the real implementation from the ideal one by calculating the fidelity \cite{nielsen2010},
\begin{equation}
    F = \abs*{\bra{\psi} U^\dagger \mathcal{E}(\ket{\psi}\bra{\psi})U \ket{\psi}}. \label{eq:fidelityeinzustand}
\end{equation}
However, this measure depends explicitly on the choice of the quantum state $\ket{\psi}$. To avoid the state dependency, one can define the \textit{average fidelity} $\bar{F}$ obtained by averaging over the fidelity of all possible quantum states,
\begin{equation}
    \bar{F}(\mathcal{E},U) = \int \bra{\psi}U^\dagger \mathcal{E}(\ket{\psi}\bra{\psi})U\ket{\psi}\mathrm{d}\psi .
\end{equation}
In Ref.~\cite{nielsen2002} it is demonstrated, that,
 in a $d$-dimensional Hilbert space, the average fidelity can be calculated by choosing a basis $U_j / \sqrt{d}$ of $d\times d$ matrices, which form an orthonormal basis with respect to the Hilbert-Schmidt scalar product,
\begin{equation}
    \bar{F}(\mathcal{E},U) = \frac{\sum_j \mathrm{tr} (UU_j^\dagger U^\dagger \mathcal{E}(U_j))  + d^2}{d^2(d+1)} . \label{eq:averagefidelity}
\end{equation}
Here, the Hilbert space is four-dimensional and spanned by the basis states $\{\ket{0}=\ket{+,+},\ket{1}=\ket{+,-},\ket{2}=\ket{-,+},\ket{3}=\ket{-,-}\}$. Hence, the orthonormal set of matrices can be chosen as $X^k Z^l$ with $k,l = 0,1,2,3$, $X\ket{j} \equiv \ket{j + 1 \mod 4}$ and $Z\ket{j} \equiv e^{ i j \pi /2} \ket{j}$.


\section{Resonator damping\label{sec:appcavitydamping}}

The resonator field $\propto(a+a^{\dagger})$ couples to an external electromagnetic environment. In order to capture the impact of this coupling to the environment on the system dynamics, we first note that, according to \eqref{eq:SWoperators}, $(a+a^{\dagger})$ transforms to the reference frame set by the SW transformation up to first order in the perturbation $V$ as,
\begin{align}
    a+&a^{\dagger}\rightarrow a+a^{\dagger}\nonumber\\
    &-\sum_{i=1}^2\frac{g_c\sin\theta^{(i)}}{\omega_c+\Omega^{(i)}}\left[\left(\ket{0}\bra{0}-\ket{1}\bra{1}+\sqrt{2}\ket{2}\bra{0}\right)\tau_+^{(i)}\right.\nonumber\\
    &\left.\hspace{64pt}+\left(\ket{0}\bra{0}-\ket{1}\bra{1}+\sqrt{2}\ket{0}\bra{2}\right)\tau_-^{(i)}\right]\nonumber\\
    &-\sum_{i=1}^2\frac{g_c\sin\theta^{(i)}}{\omega_c-\Omega^{(i)}}\left[\left(\ket{0}\bra{0}-\ket{1}\bra{1}+\sqrt{2}\ket{2}\bra{0}\right)\tau_-^{(i)}\right.\nonumber\\
    &\left.\hspace{64pt}+\left(\ket{0}\bra{0}-\ket{1}\bra{1}+\sqrt{2}\ket{0}\bra{2}\right)\tau_+^{(i)}\right]
    \nonumber\\
    &-\sum_{i=1}^2\frac{g_c\cos\theta^{(i)}}{\omega_c}\Big(2\ket{0}\bra{0}- 2\ket{1}\bra{1}\nonumber\\
    &\hspace{75pt}+\sqrt{2}\ket{2}\bra{0}+\sqrt{2}\ket{0}\bra{2}   \Big)\tau_z^{(i)}.
\end{align}
We aim at deriving a Lindblad master equation that includes the interaction of the system with the electromagnetic environment in the equation of motion for the system density matrix. 
Following the derivation of the Lindblad master equation outlined in Ref.~\cite{breuer2007}, the electromagnetic environment is traced out, and subsequently the Born-Markov and secular approximations are applied. Moreover, we assume that the environment can be approximated to be at zero temperature, such that processes increasing the system energy due to the interaction with the environment are not allowed. As a result, the impact of the environment on the system dynamics can be described by the dissipators,
\begin{align}
    &\mathcal{D}[a](\bar{\rho}), \label{eq:diss1}\\
    &\mathcal{D}\left[-\sum_{i=1}^2\frac{g_c\sin\theta^{(i)}}{\omega_c+\Omega^{(i)}}\left(\ket{1}\bra{1}-\ket{0}\bra{0}\right)\tau_-^{(i)}\right.\nonumber\\
    &\quad\,\,\left.-\sum_{i=1}^2\frac{g_c\sin\theta^{(i)}}{\omega_c-\Omega^{(i)}}\left(\ket{0}\bra{0}-\ket{1}\bra{1}\right)\tau_-^{(i)}\right](\bar{\rho}),\\
    &\mathcal{D}\left[-\sum_{i=1}^2\frac{g_c\sin\theta^{(i)}}{\omega_c+\Omega^{(i)}}\sqrt{2}\ket{0}\bra{2}\tau_-^{(i)}\right](\bar{\rho}),\\
    & \left\{\begin{matrix}
         &  \mathcal{D}\left[-\sum_{i=1}^2\frac{g_c\sin\theta^{(i)}}{\omega_c-\Omega^{(i)}}\sqrt{2}\ket{2}\bra{0}\tau_-^{(i)}\right](\bar{\rho}) \,\,\,\text{if $2\omega_c<\Omega^{(i)}$} \\
         &\\
         &  \mathcal{D}\left[-\sum_{i=1}^2\frac{g_c\sin\theta^{(i)}}{\omega_c-\Omega^{(i)}}\sqrt{2}\ket{0}\bra{2}\tau_+^{(i)}\right](\bar{\rho}) \,\,\,\text{if $2\omega_c>\Omega^{(i)}$}
    \end{matrix},
    \right.
    \\
    %
&\mathcal{D}\left[-2\sum_{i=1}^2\frac{g_c\cos\theta^{(i)}}{\omega_c}\left(\ket{0}\bra{0}-\ket{1}\bra{1}\right)\right](\bar{\rho}),\\
    &\mathcal{D}\left[-\sum_{i=1}^2\frac{g_c\cos\theta^{(i)}}{\omega_c}\left(\sqrt{2}\ket{0}\bra{2}\right)\tau_z^{(i)}\right](\bar{\rho}),    \label{eq:diss5}
\end{align}
where $\bar{\rho}$ is the density matrix in the reference frame set by the SW transformation.

Next, we note that the effective Hamiltonian $\bar{H}_d$ \eqref{eq:hameff} does not involve any coupling between subspaces with different photon numbers. Furthermore, the dissipators \eqref{eq:diss1}--\eqref{eq:diss5} do not describe processes increasing the resonator photon number in the system parameter regime characterized by $2\omega_c>\Omega^{(i)}$. These two observations lead to the following conclusion: if the system is initialized in the zero photon subspace and $2\omega_c>\Omega^{(i)}$, as considered in the main part of this paper, the system remains in the zero photon subspace during evolution. Thus, given this situation, among the dissipators \eqref{eq:diss1}--\eqref{eq:diss5} only,
\begin{align}
    &\mathcal{D}\left[\sum_{i=1}^2\left(\frac{g_c\sin\theta^{(i)}}{\omega_c+\Omega^{(i)}}-\frac{g_c\sin\theta^{(i)}}{\omega_c-\Omega^{(i)}}\right)\ket{0}\bra{0}\tau_-^{(i)}\right](\bar{\rho}), 
\end{align}
and 
\begin{align}
    \mathcal{D}\left[-2\sum_{i=1}^2\frac{g_c\cos\theta^{(i)}}{\omega_c}\ket{0}\bra{0}\tau_z^{(i)}\right](\bar{\rho}),
\end{align}
are not vanishing when acting on the system density matrix $\rho$. The above dissipators remain unchanged under the transformation to the rotating reference frame in which the Hamiltonian H \eqref{eq:hideal} is expressed.

Assuming identical qubits operated at the charge noise sweet spot ($\varepsilon=0 \rightarrow \theta=\pi/2$)  and taking into account that $\omega_c-\Omega^{(i)}\ll\omega_c+\Omega^{(i)}$ results in the dissipator used in Eq.~\eqref{eq:dglcavity} of the main text. 

\section{Qubit dephasing \label{sec:appendixdephasing}}

Considering charge qubits, charge noise, i.e., random electric fields that occur at the position of the QD, is the dominating noise source. In semiconductors, low frequency charge noise with a noise spectral density that scales (approximately) inversely proportional to frequency ($1/f$) is typically observed. Fluctuations of the level detuning $\varepsilon$ are the dominant effect of charge noise on charge qubits. According to \eqref{eq:Omega} this translates directly to a fluctuation in the charge qubit energy $\Omega$. This in turn implies that a charge qubit couples to charge noise via the operator $\tau_z$. Here, two spatially separated charge qubits are considered such that the charge noise acting on the individual qubits can be assumed to be independent and uncorrelated. In the following, identical qubits are assumed, i.e., $\theta^{(1)}=\theta^{(2)}=\theta$ and $\Omega^{(1)}=\Omega^{(2)}=\Omega$. Then, according to \eqref{eq:SWoperators}, the respective coupling operator $\tau_z^{(i)}$ transforms to the reference frame set by the SW-transformation up to second order in the perturbation $V$ as,
\begin{align}
    \tau_{z}^{(i)}&\rightarrow (1+\alpha_1 \left(\ket{1}\bra{1}+\ket{0}\bra{0}\right))\tau_{z}^{(i)}\nonumber\\
    &- 2\frac{g_c\sin\theta^{(i)}}{\omega_c + \Omega^{(i)}} \left(\ket{1}\bra{0} \tau_+^{(i)} + \ket{0}\bra{1}\tau_-^{(i)}\right) \nonumber \\
    &+2 \frac{g_c\sin\theta^{(i)}}{\omega_c -\Omega^{(i)}} \left(\ket{1}\bra{0} \tau_-^{(i)} + \ket{0}\bra{1}\tau_+^{(i)}\right)\nonumber\\
&+\alpha_2\left(1+\tau_-^{(1)}\tau_+^{(2)}+\tau_+^{(1)}\tau_-^{(2)}\right)\left(\ket{0}\bra{0}-\ket{1}\bra{1}\right)\nonumber\\
&+\alpha_3 \tau_z^{(j\neq i)}\tau_x^{(i)}\left(\ket{0}\bra{0}-\ket{1}\bra{1}\right)\nonumber\\
&+\alpha_4 \tau_x^{(i)}\left(\ket{1}\bra{1}+\ket{0}\bra{0}\right)\nonumber\\
&+ \alpha_5 (\sqrt{2}\ket{2}\bra{0} - \sqrt{2}\ket{0}\bra{2})(\tau_+^{(1)}\tau_-^{(2)} - \tau_-^{(1)}\tau_+^{(2)}) \nonumber\\
&+ \alpha_6 (\sqrt{2}\ket{0}\bra{2}\tau_-^{(1)}\tau_-^{(2)} + \sqrt{2}\ket{2}\bra{0}\tau_+^{(1)}\tau_+^{(2)}) \nonumber\\
&+ \alpha_7 (\sqrt{2}\ket{0}\bra{2}\tau_+^{(1)}\tau_+^{(2)} + \sqrt{2}\ket{2}\bra{0}\tau_-^{(1)}\tau_-^{(2)}) \nonumber\\ 
&+ \alpha_8 (\sqrt{2}\ket{0}\bra{2}\tau_+^{(2)}\frac{1+\tau_z^{(1)}}2 + \sqrt{2}\ket{2}\bra{0}\tau_-^{(2)}\frac{1+\tau_z^{(1)}}2) \nonumber\\
&+ \alpha_9 (\sqrt{2}\ket{0}\bra{2}\tau_-^{(2)}\frac{1-\tau_z^{(1)}}2 + \sqrt{2}\ket{2}\bra{0}\tau_+^{(2)}\frac{1-\tau_z^{(1)}}2) \nonumber\\
&+ \alpha_{10} (\sqrt{2}\ket{0}\bra{2}\tau_-^{(2)}\frac{1+\tau_z^{(1)}}2 + \sqrt{2}\ket{2}\bra{0}\tau_+^{(2)}\frac{1+\tau_z^{(1)}}2)\nonumber \\
&+ \alpha_{11} (\sqrt{2}\ket{0}\bra{2}\tau_+^{(2)}\frac{1-\tau_z^{(1)}}2 + \sqrt{2}\ket{2}\bra{0}\tau_-^{(2)}\frac{1-\tau_z^{(1)}}2),
\label{eq:tauzSWexpansion}
\end{align}
with 
\allowdisplaybreaks{
\begin{align}
     \alpha_1 &= - \frac{g_c^2 \sin^2\theta}{(\omega_c - \Omega)^2} - \frac{g_c^2 \sin^2\theta}{(\omega_c + \Omega)^2}, \\
    \alpha_2 &=  -\frac{g_c^2 \sin^2\theta}{(\omega_c - \Omega)^2} + \frac{g_c^2 \sin^2\theta}{(\omega_c + \Omega)^2}, \\
    \alpha_3 &=- \frac{g_c^2 \sin\theta\cos\theta}{\omega_c(\omega_c -\Omega)} + \frac{g_c^2 \sin\theta\cos\theta}{\omega_c(\omega_c + \Omega)}\\
    \alpha_4 &= + \frac{g_c^2 \sin\theta\cos\theta}{\omega_c(\omega_c -\Omega)} + \frac{g_c^2 \sin\theta\cos\theta}{\omega_c(\omega_c + \Omega)}\\
    \alpha_5 &= \frac{2 gc^2 \sin^2\theta}{\Omega^2 - \omega_c^2} \\
    \alpha_6 &= \frac{2 gc^2 \sin^2\theta}{(\omega_c + \Omega)^2} \\
    \alpha_7 &= -\frac{2 gc^2 \sin^2\theta}{(\omega_c - \Omega)^2} \\
   \alpha_8 &= -\frac{4 gc^2 \cos\theta \sin\theta}{\omega_c (2\omega_c - \Omega)} \\
    \alpha_9 &= -\frac{4 gc^2 \cos\theta \sin\theta}{\omega_c (2\omega_c + \Omega)} \\
    \alpha_{10}&= \frac{2 gc^2 \sin 2\theta}{(\omega_c + \Omega) (2\omega_c + \Omega)} \\
    \alpha_{11}&= \frac{2 gc^2 \sin 2\theta}{(\omega_c - \Omega) (2\omega_c - \Omega)}. 
\end{align}}

Now, we proceed as explained in App.~\ref{sec:appcavitydamping}. Given the low-frequency character of charge noise, only processes conserving the system energy are supported, so that charge noise is accounted for by the dissipator,
\begin{align}
\mathcal{D}&\left[(1+\alpha_1)\tau_{z}^{(i)}\right.\nonumber\\
 &\left.-\alpha_2\left(1+\tau_-^{(1)}\tau_+^{(2)}+\tau_+^{(1)}\tau_-^{(2)}\right)\left(\ket{1}\bra{1}-\ket{0}\bra{0}\right)\right](\bar{\rho}),
\end{align}
whereby $\omega_c\neq \Omega^{(i)}$ and $\Omega^{(1)}=\Omega^{(2)}$ is assumed. Applying the same reasoning as in App.~\ref{app:charge_dephasing}, the dissipator simplifies to,
\begin{align}
\mathcal{D}&\left[(1+\alpha_1)\tau_{z}^{(i)}\right.\nonumber\\
 &\left.+\alpha_2\left(1+\tau_-^{(1)}\tau_+^{(2)}+\tau_+^{(1)}\tau_-^{(2)}\right)\ket{0}\bra{0}\right](\bar{\rho}),
\end{align}
if the system is initialized in the zero photon subspace. The above dissipator remains unchanged under the transformation to the rotating reference frame in which the Hamiltonian H \eqref{eq:hideal} is expressed.

\section{Charge dephasing rate \label{app:charge_dephasing}}

In order to determine the charge dephasing rate, we first consider a general qubit Hamiltonian $\tilde{H}(\varepsilon)$ that depends on a quantity $\varepsilon$ subjected to noise. Let us now define the unitary operator $U(\varepsilon)$ that transforms $\tilde{H}(\varepsilon)$ to its eigenbasis, i.e.,
\begin{align}
H(\varepsilon)=U(\varepsilon)\tilde{H}(\varepsilon)U^{\dagger}(\varepsilon)=\frac{\omega(\varepsilon)}{2}\tau_z,
\end{align}
where $\omega(\varepsilon)/2\pi$ is the qubit transition frequency. Now, suppose that the unperturbed Hamiltonian is given for $\varepsilon=\varepsilon_0$. Then, the Hamiltonian for a general $\varepsilon$ in the eigenbasis of the unperturbed Hamiltonian reads,
\begin{align}
H(\varepsilon)=U(\varepsilon_0)\tilde{H}(\varepsilon)U^{\dagger}(\varepsilon_0). 
\end{align}
As the next step, we expand $H(\varepsilon)$ about the unperturbed value $\varepsilon=\varepsilon_0$ up to second order, 
\begin{align}
H(\varepsilon)=& H(\varepsilon_0)+\frac{\mathrm{d}H(\varepsilon)}{\mathrm{d}\varepsilon}\Bigg|_{\varepsilon_0}\delta\varepsilon+\frac{1}{2}\frac{\mathrm{d}^2H(\varepsilon)}{\mathrm{d}^2\varepsilon}\Bigg|_{\varepsilon_0}{\delta\varepsilon}^2+\mathcal{O}\left({\delta\varepsilon}^3\right)\label{eq:Hepsilonexpanded}\\
=&\frac{1}{2}\left[\left(\omega(\varepsilon_0)+\delta\omega_z\right)\tau_z+\delta\omega_x\tau_x+\delta\omega_y\tau_y\right],
\label{eq:Hepsilonexpandedpara}
\end{align}
where $\delta\varepsilon=\varepsilon-\varepsilon_0$. Equation~\eqref{eq:Hepsilonexpandedpara} follows from a decomposition of Eq.~\eqref{eq:Hepsilonexpanded} into factors multiplying the Pauli operators. A thorough analysis reveals that the parameters can be calculated as follows, 
\begin{align}
    \delta\omega_z=&\frac{\mathrm{d}\omega(\varepsilon)}{\mathrm{d}\varepsilon}\Bigg|_{\varepsilon_0}\delta\varepsilon+\frac{1}{2}\frac{\mathrm{d}^2\omega(\varepsilon)}{\mathrm{d}^2\varepsilon}\Bigg|_{\varepsilon_0}{\delta\varepsilon}^2,\\
\delta\omega_x=&\Re\left[\bra{g(\varepsilon_0)}\left(\frac{\mathrm{d}\tilde{H}(\varepsilon)}{\mathrm{d}\varepsilon}\Bigg|_{\varepsilon_0}\right)\ket{e(\varepsilon_0)}\right]\delta\varepsilon\nonumber\\
    &+\frac{1}{2}\Re\left[\bra{g(\varepsilon_0)}\left(\frac{\mathrm{d}^2\tilde{H}(\varepsilon)}{\mathrm{d}^2\varepsilon}\Bigg|_{\varepsilon_0}\right)\ket{e(\varepsilon_0)}\right]{\delta\varepsilon}^2,\\
    \delta\omega_y=&\Im\left[\bra{g(\varepsilon_0)}\left(\frac{\mathrm{d}\tilde{H}(\varepsilon)}{\mathrm{d}\varepsilon}\Bigg|_{\varepsilon_0}\right)\ket{e(\varepsilon_0)}\right]\delta\varepsilon\nonumber\\
    &+\frac{1}{2}\Im\left[\bra{g(\varepsilon_0)}\left(\frac{\mathrm{d}^2\tilde{H}(\varepsilon)}{\mathrm{d}^2\varepsilon}\Bigg|_{\varepsilon_0}\right)\ket{e(\varepsilon_0)}\right]{\delta\varepsilon}^2,
\end{align}
with $\ket{g(\varepsilon_0)}$ and $\ket{e(\varepsilon_0)}$ the ground state and the excited state of the unperturbed Hamiltonian $\tilde{H}(\varepsilon_0)$, respectively. 

Given a Hamiltonian of the form \eqref{eq:Hepsilonexpandedpara}, assuming Gaussian distributed noise with zero mean and a power spectral density $S_{\varepsilon}(\tilde{\omega})=A_{\varepsilon}/  |\tilde{\omega}|$ with constant $A_{\varepsilon}$, as typically observed for charge noise, a Ramsey free-induction decay sequence will lead to a Gaussian decay $\propto\exp\left[-(t/T_{\phi})^2\right]$ with dephasing time \cite{russ2015},
\begin{align}
    T_{\phi}=&\left[\frac{1}{2}\left(\frac{\mathrm{d}\omega(\varepsilon)}{\mathrm{d}\varepsilon}\Bigg|_{\varepsilon_0}\right)^2A_{\varepsilon}\log(r)\right.\nonumber\\
    &\,\,\,\left.+\frac{1}{4}\left(\frac{\mathrm{d}^2\omega(\varepsilon)}{\mathrm{d}^2\varepsilon}\Bigg|_{\varepsilon_0}\right)^2A^2_{\varepsilon}\log^2(r)\right]^{-\frac{1}{2}},
\end{align}
with $r=\frac{\omega_U}{\omega_R}$, where $\omega_U$ and $\omega_R$ are cutoff frequencies ${\omega_U\leq\tilde{\omega}\leq\omega_R}$ that are not relevant for the reasoning in this manuscript. 

Here, the qubit Hamiltonian $\tilde{H}(\varepsilon)$ is given by the DQD Hamiltonian $\tilde{H}_{\mathrm{DQD}}$ (Eq.~\eqref{eq:HDQD}). This Hamiltonian depends on two parameters, $\varepsilon$ and $t_c$, both susceptible to charge noise. However, the effect of charge noise on the tunnel coupling strength $t_c$ is often negligible compared to that on the level detuning $\varepsilon$ \cite{benito2019a}. Thus, only the level detuning is considered as a fluctuating parameter in the following. 
Following the above reasoning, we find
\allowdisplaybreaks{
\begin{align}
    \omega(\varepsilon)&=\sqrt{\varepsilon^2+(2t_c)^2},\\
    \delta\omega_z&=\frac{\varepsilon_0}{\omega(\varepsilon_0)}\delta\varepsilon+\frac{1}{2}\left(\frac{1}{\omega(\varepsilon_0)}-\frac{\varepsilon_0^2}{\omega(\varepsilon)^3}\right){\delta\varepsilon}^2,
    \label{eq:deltaomegazDQD}\\
    \delta\omega_x&=-\frac{t_c}{\omega(\varepsilon_0)}\delta\varepsilon,\\
    \delta\omega_y&=0.
\end{align}}
Equation~\eqref{eq:deltaomegazDQD} shows that the first-order fluctuations of the qubit energy splitting can be suppressed by operating the system at the charge noise sweet spot characterized by $\varepsilon_0=0$. Setting the system to this particular configuration, one has 
\begin{align}
&\omega(\varepsilon_0=0)=2t_c,\\
&    \delta\omega_z=\frac{1}{2}\frac{1}{2t_c}{\delta\varepsilon}^2,\\
    &\delta\omega_x=-\frac{1}{2}\delta\varepsilon,
\end{align}
and 
\begin{align}
    T_{\phi}=2\pi\left[\frac{1}{4}\left(\frac{1}{2t_c}\right)^2A_{\varepsilon}^2\log^2(r)\right]^{-\frac{1}{2}}.
    \label{eq:dephasingtimeDQD}
\end{align}
 Given the above expression for the dephasing time it follows that the dephasing rate,
\begin{align}
    \gamma_{\phi}=\frac{2\pi}{T_{\phi}}\propto\frac{1}{t_c},
    \label{eq:gammaphi}
\end{align}
is inversely proportional to the tunnel coupling strength $t_c$. In a recent experiment, a charge dephasing rate of $\gamma_{\phi}/2\pi=9.9\,\mathrm{MHz}$ with the tunnel coupling strength set to  $2t_c/h=19.2\,\mathrm{GHz}$ was observed \cite{yu2023}. According to Eq.~\eqref{eq:gammaphi}, the reported values allow us to infer the proportionality factor $C$ between the charge dephasing rate $\gamma_{\phi}$ and the inverse tunnel coupling strength $1/t_c$,
\begin{align}
    \frac{\gamma_{\phi}(t_c)}{2\pi}=\frac{C}{t_c},
    \label{eq:gammaphioftc}
\end{align}
with $C=393\,\mu\mathrm{eV}\times\mathrm{MHz}$. The dephasing rate given in Eq.~\eqref{eq:gammaphioftc} is employed in the calculations in Sec.~\ref{sec:charge_qubit_dephasing} of the main text. We note that a master equation  of the form, \begin{align}
    \dot{\rho}^{\tau}(t)=-i\left[t_c\tau_z,\rho^{\tau}(t)\right]+\frac{\gamma_{\phi}}{2}\mathcal{D}[\tau_z]\rho^{\tau}(t),
 \end{align} 
 where the dephasing rate $\gamma_{\phi}=2\pi/T_\phi$ describes an exponential decay $\propto \exp\left[-t/T_{\phi}\right]$ of the coherences of the charge qubit density matrix $\rho^{\tau}(t)$ of an individual charge qubit, while the discussion in this Appendix predicts a Gaussian decay. We point out that $\exp\left[-t/T_{\phi}\right]\leq\exp\left[-(t/T_{\phi})^2\right]$ in the time domain $0\leq t\leq T_{\phi}$. Thus, on this timescale, the master equation description overestimates the impact of charge noise on charge dephasing. As the gate times in the main text are shorter than the charge dephasing time, it is guaranteed that the impact of charge noise is not underestimated in our calculations.

\bibliography{library}

\end{document}